\documentclass[preprint,12pt]{elsarticle}
\usepackage{ifpdf}
\usepackage{graphicx}
\usepackage{amssymb}
\usepackage{amsmath}
\makeatletter
\def\ps@pprintTitle{%
 \let\@oddhead\@empty
 \let\@evenhead\@empty
 \def\@oddfoot{}%
 \let\@evenfoot\@oddfoot}
\makeatother
\begin{document}

\begin{frontmatter} 

\title{Order preserving contact transformations and dynamical symmetries of scalar and coupled Riccati and Abel chains}
\author[gla]{R. Gladwin Pradeep}
\author[vkc]{V. K. Chandrasekar}
\author[bdu]{R. Mohanasubha}
\author[bdu]{M. Senthilvelan\corref{cor1}}
\ead{velan@cnld.bdu.ac.in}

\author[bdu]{M. Lakshmanan}
\address[gla]{Department of Physics, KCG College of Technology, Karapakkam, Chennai - 600 097, India}
\address[vkc]{Centre for Nonlinear Science and Engineering, School of Electrical and Electronics Engineering, SASTRA University, Thanjavur - 613 401, India}
\address[bdu]{Centre for Nonlinear Dynamics, School of Physics, Bharathidasan University, Tiruchirappalli - 620 024, India}
\cortext[cor1]{Corresponding author} 





\begin{abstract}
We identify contact transformations which linearize the given equations in the Riccati and Abel chains of nonlinear scalar and coupled ordinary differential equations to the same order. The identified contact transformations are not of Cole-Hopf type and are \emph {new} to the literature. The linearization of Abel chain of equations is also demonstrated explicitly for the first time. The contact transformations can be utilized to derive dynamical symmetries of the associated nonlinear ODEs. The wider applicability of identifying this type of contact transformations and the method of deriving dynamical symmetries by using them is illustrated through two dimensional generalizations of the Riccati and Abel chains as well.
\end{abstract}

\begin{keyword}
Contact transformation, Linearization,  Dynamical symmetries, Riccati and Abel chains.
\end{keyword}

\end{frontmatter}

\section{Introduction}

Solving a nonlinear ordinary differential equation (ODE) by transforming it to a linear ODE is one of the classical methods of finding solutions in the theory of differential equations.  Such a study is called linearization/equivalence problem \cite{olver}. The linearization can be achieved by identifying a suitable transformation. The transformation may be a point transformation \cite{steeb:book:93:01}, a contact transformation \cite{19}, a Sundman transformation \cite{duarte:94:01,berkovich,euler:2003}, a nonlocal transformation \cite{chandrasekar:06:01:jpa,gladwin:jpa:2011,gladwin:jmp:2010}, or a generalized linearizing transformation \cite{gladwin:jpa:2006}, to name a few. The transformation can be identified by analyzing the structure of the ODE. As far as second order nonlinear ODEs are concerned, it has been shown that the general form of the nonlinear ODE which can be transformed to the free particle equation by an invertible point transformation should be at the most cubic in the first derivative and its coefficients have to satisfy two conditions involving their partial derivatives \cite{mahomed:1989a}. Subsequently it has been proved that any second order nonlinear ODE which admits maximal Lie point symmetries (eight) also satisfies the lineariziability criteria \cite{mahomed:1989a, imb}. A systematic procedure is also available to construct invertible point transformations from the Lie point symmetries \cite{mahomed:1985}. In other words, one can find a connection between invertible point transformations and Lie  point symmetries. Linearization of a second order nonlinear ODE by Sundman transformation was studied by Durate et al. \cite{duarte:94:01}. Unlike the case of invertible point transformation, in the present case, the new independent variable is considered in a nonlocal form  \cite{chandrasekar:06:01:jpa}. The connection between Sundman symmetries and Sundman transformation was studied in detail by Euler et al \cite{euler:2003}. Recently, three of the present authors have introduced a generalized linearizing transformation in which the new independent variable is not only nonlocal in form but also involves derivatives \cite{gladwin:jpa:2006}. However, the connection between generalized linearizing transformations and symmetries is not yet known. Apart from the above one can also find other linearizing transformations, for example, nonlocal transformations that can linearize a class of equations, see for example \cite{abraham:1992,abraham-shrauner:4809,abraham,govinder:1995} and references therein. Again the interrelation between nonlocal transformations and symmetries is yet to be proved in general.



In this paper, we report a new kind of contact transformation which linearizes the given equation in the same order. From the literature it is known that the contact transformation linearizes a given ODE to a new ODE which is one order higher than the original one. The well known example is the modified Emden equation (MEE), $\ddot{x}+3 x \dot{x}+x^3=0$. The contact transformation $x=\frac{\dot{w}}{w}$ linearizes the above equation to an equation of third order, that is $\dddot{w}=0$. On the other hand, as we show below, the above Li$\acute{e}$nard type nonlinear oscillator equation can also be linearized to the free particle equation $\frac{d^2u}{dt^2}=0$ through the contact transformation $u=\frac{x}{2\dot{x}+x^2}$. As one witnesses, the transformation does not increase the order of the underlying ODE. Unlike the known contact transformations (which are identified by an ad-hoc way) the ones which we report here can be derived in an algorithmic manner. An immediate consequence of finding this type of contact transformations is that one can look for dynamical symmetries associated with the underlying nonlinear ODEs. The latter result is significant since the dynamical symmetries are usually difficult to derive and no systematic method exists in the literature to explore them. This is because once the infinitesimal generators in the Lie symmetry analysis are allowed to involve derivative terms then one can no longer determine the complete symmetry group. Dynamical symmetries of ODEs can be regarded as the transformations of the set of first integrals \cite{sir1,hydon}. Deviating from the conventional approach of solving the determining equations arising from the associated invariance condition, in this paper, we develop a simple and straightforward algorithm to derive dynamical symmetries associated with the given nonlinear ODE belonging to the Riccati and Abel chains of ODEs. Our work also reveals the hidden connection between contact transformations and dynamical symmetries. It is known that the MEE equation is the second member of the Riccati chain (see below), and the third member in this chain is a third order nonlinear ODE and the fourth member is a fourth order nonlinear ODE and so on \cite{gladwin:jmp:2010}. The procedure which we develop here can also be extended to any of the equations in this chain. We also bring further light into this theory by considering the other chains, namely Abel chain and coupled Riccati and Abel chains. In all these chains the same methodology can be adopted to obtain the linearizing contact transformations and the dynamical symmetries of the given equation.

We note here that the linearization of the Riccati chain through the Cole-Hopf transformation is known. However, as far as the authors' knowledge goes, the linearization of any of the equations in the Abel chain has not been demonstrated so far. In this paper, we show the exact linearization of the equations admitted by the Abel chain through contact transformation. 
The integrable scalar Riccati and Abel chains are of the forms \cite{carinena:2009,carinena,chandrasekar:royal1,euler:2009,euler:2007,gladwin:jmp:2010},
\begin{eqnarray}
\left(\frac{d}{dt}+kx\right)^mx=0,\qquad m=1,2,3,\ldots\label{chap7-riccatichain}
\end{eqnarray}
and 
\begin{eqnarray}
\left(\frac{d}{dt}+kx^2\right)^mx=0,\qquad m=1,2,3,\ldots\label{chap7-abelchain}
\end{eqnarray} respectively. Interestingly one may also generalize the above differential equations to 
\begin{eqnarray}
\left(\frac{d}{dt}+f(x,t)\right)^mg(x,t)=0,\qquad m=1,2,3,\ldots.\label{chap7-gen-riccati}
\end{eqnarray}
The above two chains possess several interesting geometric structures and certain common mathematical properties, see for example \cite{carinena:2009,bruzon}. The results which we present here will further illuminate the shared mathematical properties of these two chains.

The methodology which we adopt to derive the aforementioned results is the following. In one of our earlier works \cite{gladwin:jmp:2010} we have shown that the Riccati and Abel chains can be transformed to the linear equation 
\begin{eqnarray}
\frac{d^mu}{dt^m}=0,~~m=2,3,4...\label{chap7-linear1}
\end{eqnarray}
through the nonlocal transformation 
\begin{equation}
 u=g(t,x) e^{\int f(t,x) dt}.\label{gh1df}
\end{equation}
For example, in the case of MEE we have $f=g=x$ and the transformed equation is just $\ddot{u}=0$. The general solution can be derived by solving the first order ODE \cite{gladwin:jpa:2011}, that is
\begin{eqnarray}
\frac{\dot{u}} {u}=\frac{\dot{g}+gf} {g},\label{identity}
\end{eqnarray} 
which can be obtained from (\ref{gh1df}) itself. Since the left hand side of Eq.(\ref{identity}) is a function of $t$ and is known from the linearized equation, we can rewrite the above equation as an equation for $\dot{x}$. Integrating the latter equation we can obtain the general solution of the nonlinear ODE. For more details one may refer to \cite{chandrasekar:06:01:jpa}. After analyzing the expression (\ref{identity}) carefully, we found that the contact transformation can also be captured from the identity (\ref{identity}) itself. To extract the contact transformation we split Eq.(\ref{identity}) into two separate expressions, one for $u$ and the other for $\dot{u}$ with an unknown function in them (see Eq.(\ref{chap7-eq-decompose}) given below). We then determine the explicit form of the unknown function which in turn provides the necessary contact transformation for the given equation. We will also illustrate the method of finding the general solution of the nonlinear ODE from the solution of the linearized equation. We then proceed to obtain the dynamical symmetries of the associated nonlinear ODE. Inverting the expressions $u$ and $\dot{u}$ for $x$ and $\dot{x}$, we obtain the necessary contact transformation ($x$ and $\dot{x}$ in terms of $u$ and $\dot{u}$). Considering the Lie point symmetries of the linearized equation and properly rewriting the variables (of linear ODE) that appear in the vector fields in terms of the original variables (of nonlinear ODE) we can obtain the dynamical symmetries of the given equation. The wider applicability of this method is demonstrated by considering the coupled Riccati and Abel chains of equations.

The plan of the paper is as follows: In section 2, we present the general theory to find the nonlocal transformation which reduces the complicated scalar nonlinear ODEs into linear ODEs. In section 3, we construct linearizing contact transformations by considering one equation each from the Riccati and Abel chains respectively. In section 4, we present the method of deriving dynamical symmetries from contact transformation for the given equation. To illustrate the algorithm we again consider one equation each from the Riccati and Abel chains. In Sec.5, we extend the theory to two coupled Riccati and Abel chains. Here also we consider one equation each from the Riccati and Abel chains and construct their linearizing contact transformations and their associated dynamical symmetries. In Sec.6, we present our conclusions. In Appendix A, we briefly demonstrate the method of identifying linearizing contact transformations and dynamical symmetries of third order nonlinear ODEs. Finally, in Appendix B, we prove that the derived dynamical symmetries satisfy the invariance condition. 

\section{General Theory}

In this section, we present the method of constructing contact transformations for the Riccati and Abel chains. For this purpose, we decompose the identity (\ref{identity}) into two equations by introducing an arbitrary function $h(t,x,\mathbf{x}^{(m-1)})$, that is
\begin{subequations}
\begin{eqnarray}
\addtocounter{equation}{-1}
\label{chap7-eq-decompose}
\addtocounter{equation}{1}
&&u=\frac{g(x,t)}{h(t,x,\mathbf{x}^{(m-1)})}\label{chap7-eq-decompose311}\\
&&\hspace{-1.5cm}\mbox{and}\nonumber\\
&&\dot{u}=\frac{\dot{g}(x,t)+g(x,t)f(x,t)}{h(t,x,\mathbf{x}^{(m-1)})},\quad x^{(n)}=\frac{d^nx}{dt^n},\label{chap7-eq-decompose411}
\end{eqnarray}
\end{subequations}
where $\mathbf{x}^{(m-1)}=\{x^{(1)},x^{(2)},\ldots,x^{(m-1)}\}$ with $x^{(n)}=\frac{d^nx}{dt^n}$. Differentiating Eq. (\ref{chap7-eq-decompose311}) with respect to `$t$' and substituting it into Eq. (\ref{chap7-eq-decompose411}) we obtain the following first order ODE for the unknown function $h(t,x,\mathbf{x}^{(m-1)})$:
\begin{eqnarray}
\frac{dh(t,x,\mathbf{x}^{(m-1)})}{dt}+f(x,t)h(t,x,\mathbf{x}^{(m-1)})=0.\label{chap7-eq-h211}
\end{eqnarray}
By solving Eq. (\ref{chap7-eq-h211}) we can obtain the function $h(t,x,\mathbf{x}^{(m-1)})$. 
To determine this function we consider Eq. (\ref{chap7-eq-h211}) in the form
\begin{eqnarray}
\frac{D\left[h(t,x,\mathbf{x}^{(m-1)})\right]}{h(t,x,\mathbf{x}^{(m-1)})}=-f(x,t),\qquad D=\frac{d}{dt}.\label{chap7-eq-h311}
\end{eqnarray}
Now rewriting the general form of the nonlinear ODE (\ref{chap7-gen-riccati}) as
\begin{equation}
\bigg(\frac{d} {dt}+f(x,t)\bigg)^{m-1}\bigg(\frac{d} {dt}+f(x,t)\bigg)g(x,t)=0\label{rela11}
\end{equation}
and comparing (\ref{chap7-eq-h211}) with (\ref{rela11}), we find 
\begin{eqnarray}
h(t,x,\mathbf{x}^{(m-1)})=\left(\frac{d}{dt}+f(x,t)\right)^{m-1}g(x,t),\quad m> 1.\label{hsol111}
\end{eqnarray}
Substituting the above form of $h(t,x,\mathbf{x}^{(m-1)})$ into (\ref{chap7-eq-decompose}), we get
\begin{subequations}
\begin{eqnarray}
\addtocounter{equation}{-1}
\label{chap7-gen-cont111}
\addtocounter{equation}{1}
&&u=\frac{g(x,t)}{\left(\frac{d}{dt}+f(x,t)\right)^{m-1}g(x,t)},\label{chap7-gen-cont1111}\\
&&\dot{u}=\frac{\dot{g}(x,t)+g(x,t)f(x,t)}{\left(\frac{d}{dt}+f(x,t)\right)^{m-1}g(x,t)}\label{chap7-gen-cont1112}.
\end{eqnarray}
\end{subequations}

However, we observe that the transformation (\ref{chap7-gen-cont111}) does not meet our requirement completely. For example, let us consider the case $m=2$. Upon restricting $m=2$ in the above expressions (\ref{chap7-gen-cont1111}) and (\ref{chap7-gen-cont1112}), one may notice that the denominator of (\ref{chap7-gen-cont1112}) matches with the numerator which in turn yields $\dot{u}=1$. In fact, this inadequacy persists at all orders. To demonstrate this, let us extend the above analysis to an arbitrary order. In this case, we find the following expression for the variable $\frac{d^{n-1}u} {dt^{n-1}}$, that is
\small
\begin{eqnarray}
\frac{d^{(n-1)}u}{dt^{(n-1)}}=\frac{\left(\frac{d}{dt}+f(x,t)\right)^{n-1}g(x,t)}{h(t,x,\mathbf{x^{(m-1)}})}=\frac{\left(\frac{d}{dt}+f(x,t)\right)^{n-1}g(x,t)}{\left(\frac{d}{dt}+f(x,t)\right)^{m-1}g(x,t)},\quad n=1,2,\ldots, m.\label{gen-contact1}
\end{eqnarray}
\normalsize
One may observe that the highest derivative of $u$ becomes one when $n=m$. In other words, the obtained forms of $u$ and their derivatives do constitute a partial transformation only. 

By carefully examining the origin of this obstacle we observe that the function $h(t,x,\mathbf{x}^{(m-1)})$ given in Eq. (\ref{hsol111}) is fixed by comparing Eqs.(\ref{chap7-eq-h211}) and (\ref{rela11}). 
 To associate the unknown function $h(t,x,\dot{x})$ with the given nonlinear ODE in order to get nontrivial results, we consider $h(t,x,\mathbf{x}^{(m-1)})$ to be of the form
\begin{eqnarray}
h(t,x,\mathbf{x}^{(m-1)})=F(I)\left(\frac{d}{dt}+f(x,t)\right)^{m-1}g(x,t),\quad m> 1,\label{hsol211}
\end{eqnarray}
where $I$ is the first integral of Eq. (\ref{rela11}) and $F$ is an arbitrary function of $I$. Note that since $I$ is an integral of motion, multiplying Eq. (\ref{hsol111}) by a constant does not make any change in Eq.(\ref{chap7-eq-h311}). Substituting Eq. (\ref{hsol211}) into (\ref{chap7-eq-h211}) we get $\frac{dF(I)}{dt}=0$.
From this general form of $h(t,x,\mathbf{x}^{(m-1)})$ we can get the necessary contact transformation. 

For simplicity, we illustrate the above theory by considering the second order ODEs in the Riccati and Abel chains. 

\section{Contact transformations} 
For illustration purpose, in the following, we construct linearizing contact transformations by considering one equation each from the Riccati and Abel chains, Eqs.(\ref{chap7-riccatichain}) and (\ref{chap7-abelchain}) respectively. 
\subsection{An example from Riccati chain}
To begin, we consider the second member in the Riccati chain, namely
\begin{eqnarray}
\ddot{x}+3x\dot{x}+x^3=0.\label{chap7-riccatichain1}
\end{eqnarray}
This equation can be obtained by fixing $f(x,t)=g(x,t)=x$ and $m=2$ in Eq. (\ref{rela11}). To identify the contact transformation of this equation we have to find the function $h(t,x,\dot{x})$. This can be determined by solving Eq.(\ref{chap7-eq-h211}). This equation resembles the determining equation for the Darboux polynomial \cite{dar1} and the integral associated with this equation can be determined in an algorithmic way. In fact, the ratio of two particular solutions of Eq.(\ref{chap7-eq-h211}) defines a first integral \cite{dar1}. Since $f$ and $g$ are fixed now, one solution can be readily found from Eq.(\ref{hsol111}) itself. For example, in the case of Eq.(\ref{chap7-riccatichain1}), since $f=g=x$, Eq.(\ref{hsol111}) straightforwardly provides $h_1=\dot{x}+x^2$, which can be treated as one particular solution. By solving (\ref{chap7-eq-h211}) we can find a second particular solution as $h_2=\sqrt{2\dot{x}+x^2}$. The ratio of these two expressions gives the first integral of (\ref{chap7-riccatichain1}), that is
\begin{eqnarray}
I=\frac{h_2} {h_1}=\frac{\sqrt{2\dot{x}+x^2}}{\dot{x}+x^2},~~~~\frac{dI}{dt}=0.\label{integral1}
\end{eqnarray}
Without loss of generality, we have assumed $F(I)=I$ throughout this paper. With this form of integral $I$, the arbitrary function $h(t,x,\dot{x})$ becomes (vide Eq.(\ref{hsol211}))
\begin{eqnarray}
 h(t,x,\dot{x})=\sqrt{2\dot{x}+x^2}.\label{hfi_sca}
\end{eqnarray}

Substituting $f=g=x$ and the function $h=\sqrt{2\dot{x}+x^2}$ into Eqs.(\ref{chap7-eq-decompose311}) and (\ref{chap7-eq-decompose411}), we can get the contact transformation as
\begin{eqnarray}
u=\frac{x}{\sqrt{2\dot{x}+x^2}},~~
\dot{u}=\frac{(\dot{x}+x^2)}{\sqrt{2\dot{x}+x^2}}.\label{conct2}
\end{eqnarray}
Rewriting Eq.(\ref{conct2}) for $x$ and $\dot{x}$, we find
\begin{eqnarray}
x=\frac{2u\dot{u}}{1+u^2},\qquad\dot{x}=\frac{2\dot{u}^2(1-u^2)}{(1+u^2)^2}.\label{contscx}
\end{eqnarray}
One can straightforwardly verify that the contact transformation $u=\frac{x}{\sqrt{2\dot{x}+x^2}}$ linearizes the nonlinear ODE (\ref{chap7-riccatichain1}) into the free particle equation $\ddot{u}=0$ (which is of the same order as that of the original equation (\ref{chap7-riccatichain1})). 

Interestingly, from the solution of the linear equation $\ddot{u}=0$, we can also derive the solution of the nonlinear ODE (\ref{chap7-riccatichain1}) as follows. Substituting the solution $u$ of the free particle equation into (\ref{conct2}) and rewriting it we obtain a first order ODE, $(\dot{x}+x^2)(t+I_1)-x=0$. Integrating this equation we can get the general solution of the nonlinear ODE (\ref{chap7-riccatichain1}) in the form
\begin{equation}
x(t)=\frac{I_1+t}{\frac{t^2}{2}+I_1t+I_2},
\end{equation}
where $I_1$ and $I_2$ are arbitrary constants. 
\subsection{An example from Abel chain}
Now we consider the second equation in the Abel chain and construct the contact transformation of it. The underlying ODE reads \cite{gladwin:jmp:2010}
\begin{eqnarray}
\ddot{x}+4x^2\dot{x}+x^5=0,\label{chap7-abelchain1}
\end{eqnarray}
which arises by fixing  $f(x,t)=x^2, g(x,t)=x$ and $m=2$ in (\ref{rela11}).  As before, we can fix the function $h(t,x,\dot{x})$ as follows. With $f=x^2,g=x$, Eq.(\ref{hsol111}) furnishes one particular solution $h_1=\dot{x}+x^3$. Upon solving Eq.(\ref{chap7-eq-h211}) with $f=x^2$, we find
\begin{eqnarray}
h_2=\frac{\sqrt{2}(\dot{x}+x^3)^{\frac{5} {2}}} {\sqrt{3 \dot{x}+x^3}}.\label{integral1}
\end{eqnarray}
From these two expressions, we can deduce $I$ to be of the form
\begin{eqnarray}
 I=\frac{h_1} {h_2}=\frac{\sqrt{3\dot{x}+x^3}} {\sqrt{2}(\dot{x}+x^3)^{\frac{3} {2}}}.
\end{eqnarray}
With the above form of $I$, we can get the expression for $h$ as
\begin{equation}
h=\frac{\sqrt{3\dot{x}+x^3}} {\sqrt{2}\sqrt{\dot{x}+x^3}}.
\end{equation}
Substituting $f=x^2,~~g=x$ and the above expression for $h$ into Eqs.(\ref{chap7-eq-decompose311}) and (\ref{chap7-eq-decompose411}), we can get the contact transformation in the form
\begin{equation}
u=\frac{\sqrt{2}x\sqrt{\dot{x}+x^3}} {\sqrt{3\dot{x}+x^3}},~~~\dot{u}=\frac{\sqrt{2}(\dot{x}+x^3)^{\frac{3} {2}}} {\sqrt{3\dot{x}+x^3}}.\label{abecon}
\end{equation}
From (\ref{abecon}), we find
\begin{eqnarray}
  x=\frac{\sqrt{\frac{3}{2} \dot{u}}u}{\sqrt{u^3+\dot{u}}},~~\dot{x}=\frac{\sqrt{\frac{3}{2}}\dot{u}^{\frac{3}{2}}(2\dot{u}-u^3)} {2(\dot{u}+u^3)^{\frac{3}{2}}}.\label{abecond}
\end{eqnarray}
One can again straightforwardly verify that the contact transformation (\ref{abecon}) linearizes the nonlinear ODE (\ref{chap7-abelchain1}) to $\ddot{u}=0$, which is of the same order as (\ref{chap7-abelchain1}). Using the contact transformation (\ref{abecon}) we can find the general solution of the second order Abel equation as well. Substituting the solution $u$ of the free particle equation into (\ref{abecon}) and rewriting it, we get the following first order ODE, namely $(\dot{x}+x^3)(t+I_1)-x=0.$ Integrating the latter we can get the general solution of Eq.(\ref{chap7-abelchain1}) in the form 
\begin{equation}
x(t)=\frac{\sqrt{3}(I_1+t)}{\sqrt{2(I_1+t)^3}+3I_2},
\end{equation}
where $I_1$ and $I_2$ are two arbitrary constants. 

One can also extend the above procedure to higher order equations in the Riccati and Abel chains and determine their contact transformations. In Appendix A, we demonstrate this application to a third order ODE.





\section{Dynamical symmetries}
In this section, we present the method of deriving dynamical symmetries from contact transformations for the given equation.  For this purpose we again consider Eqs.(\ref{chap7-riccatichain1}) and (\ref{chap7-abelchain1}).
\subsection{Algorithm}
The nonlinear ODE (\ref{chap7-riccatichain1}) admits eight Lie point symmetries \cite{pandey:102701} and is linearizable through point transformation.  Here we report the dynamical symmetries of this equation and Eq.(\ref{chap7-abelchain1}). To derive these symmetries we make use of the contact transformations (vide Eqs.(\ref{conct2}) and (\ref{abecon})) obtained earlier.

Let $\xi$ and $\eta$ be the are the coefficient functions of the associated generator of an infinitesimal transformations, that is $u'=u+\epsilon \eta(t,u)$ and $t'=t+\epsilon \xi(t,u)$, $\epsilon<<1$, of the linear ODE.  The symmetry vector field associated with the infinitesimal transformations is then given by
\begin{eqnarray}            
\Lambda=\xi(t,u) \frac{\partial}{\partial t}
+\eta(t,u) \frac{\partial}{\partial u}
\label {sym01}
\end{eqnarray}
whose first prolongation is
\begin{eqnarray}            
\Lambda^{1}=\xi \frac{\partial}{\partial t}
+\eta \frac{\partial}{\partial u}+(\dot{\eta}
-\dot{u}\dot{\xi})\frac{\partial}{\partial \dot{u}},
\label {sym02}
\end{eqnarray}
where dot represents the total derivative with respect to $t$ ($~\dot{} =\frac{d} {dt}$).

Let us designate the general symmetry vector field and its first prolongation of the nonlinear ODEs (\ref{chap7-riccatichain1}) and (\ref{chap7-abelchain1}), respectively, as
\begin{eqnarray}            
\Omega=\lambda \frac{\partial}{\partial t}
+\mu \frac{\partial}{\partial x},
\label {sym03}
\end{eqnarray}
and
\begin{eqnarray}            
\Omega^{1}=\lambda \frac{\partial}{\partial t}
+\mu \frac{\partial}{\partial x}+(\dot{\mu}
-\dot{x}\dot{\lambda})\frac{\partial}{\partial \dot{x}}.
\label {sym04}
\end{eqnarray}
In the above $\lambda$ and $\mu$ are the infinitesimal symmetries of the given nonlinear ODE, for example (\ref{chap7-riccatichain1}) or (\ref{chap7-abelchain1}). If $\lambda$ and $\mu$ are restricted to the variables $t$ and $x$ only, then they become the Lie point symmetries of the given equation. Here we consider the function $\lambda$ and $\mu$ to be not only functions of $t$ and $x$ but also a function of $\dot{x}$ as well. The associated symmetries are called dynamical symmetries and they can be utilized to establish the integrability of the given nonlinear ODE  \cite{gladwin:jmp:2010}. As we have pointed in the introduction, unlike the Lie point symmetries which can be determined in an algorithmic way, there exists no systematic procedure to derive the dynamical symmetries. 

To investigate the dynamical symmetries of the given nonlinear ODE we adopt the following methodology. We first identify the contact transformation between the variables of the nonlinear ODE and the associated linear ODE (vide Eq. (\ref{conct2}) or (\ref{abecon})). Using this transformation we can express the partial derivatives that appear in the vector field (\ref{sym02}) in terms of the variables of the nonlinear ODE and then compare the resultant expression with the vector field (\ref{sym04}) which in turn provides the dynamical symmetries for the nonlinear ODE. In the following, we demonstrate this explicitly.

From the contact transformation (for example Eq. (\ref{conct2}) or (\ref{abecon})), we can extract the following differential identities:
\begin{eqnarray}
&&\frac{\partial }{\partial u}=\frac{\partial x}{\partial u}\frac{\partial}{\partial x}+\frac{\partial \dot{x}}{\partial u}\frac{\partial}{\partial \dot{x}},~~~\frac{\partial }{\partial \dot{u}}=\frac{\partial x}{\partial \dot{u}}\frac{\partial}{\partial x}+\frac{\partial \dot{x}}{\partial \dot{u}}\frac{\partial}{\partial \dot{x}}.\label{b111}
\end{eqnarray}
Rewriting the first prolongation $\Lambda^1$ (vide Eq.(\ref{sym02})) by using the above expressions, we get
\begin{eqnarray}
\Lambda^1=\xi\frac{\partial}{\partial t}+\left(\eta\frac{\partial x}{\partial u}+(\dot{\eta}-\dot{u}\dot{\xi})\frac{\partial x}{\partial \dot{u}}\right)\frac{\partial}{\partial x}+
\left(\eta\frac{\partial \dot{x}}{\partial u}+(\dot{\eta}-\dot{u}\dot{\xi})\frac{\partial \dot{x}}{\partial \dot{u}}\right)\frac{\partial}{\partial \dot{x}}.\label{lambda1}
\end{eqnarray}
Comparing the prolongation vectors (\ref{lambda1}) and (\ref{sym04}), we find
\begin{eqnarray}
\lambda=\xi,\qquad
\mu=\eta\frac{\partial x}{\partial u}+(\dot{\eta}-\dot{u}\dot{\xi})\frac{\partial x}{\partial \dot{u}}.\label{arbitrary-sym1}
\end{eqnarray}
One can check that $\dot{\mu}-\dot{x}\dot{\xi}=\eta\frac{\partial \dot{x}}{\partial u}+(\dot{\eta}-\dot{u}\dot{\xi})\frac{\partial \dot{x}}{\partial \dot{u}}$. Since $x$ is known in terms of $u$ and $\dot{u}$, Eq.(\ref{arbitrary-sym1}) can be evaluated straightforwardly. In other words the infinitesimal symmetries $\lambda$ and $\mu$ are now found in terms of $x$ and $\dot{x}$. Since there exists eight infinitesimal Lie point symmetries for the free particle equation, we may consider each one of them separately, and by substituting them into (\ref{arbitrary-sym1}) we can identify the dynamical symmetries associated with the nonlinear ODE. 

\subsection{Dynamical symmetries of Eq.(\ref{chap7-riccatichain1})}
In the following, we apply the above algorithm to Eq.(\ref{chap7-riccatichain1}) and derive the dynamical symmetries of it. Recalling the transformation (\ref{conct2}) and substituting the expressions
\begin{eqnarray}
\frac{\partial x} {\partial u}=\frac{\dot{x}\sqrt{2\dot{x}+x^2}} {\dot{x}+x^2},~~
\frac{\partial x} {\partial \dot{u}}=\frac{x\sqrt{2\dot{x}+x^2}} {\dot{x}+x^2}.
\end{eqnarray}
into (\ref{arbitrary-sym1}), we find
\begin{eqnarray}
\lambda=\xi,\qquad
\mu=\frac{\sqrt{2\dot{x}+x^2}}{\dot{x}+x^2}\left(\eta\dot{x}+\dot{\eta}x\right)-\dot{\xi}x,\label{arbitrary-sym}
\end{eqnarray}
where $\xi$ and $\eta$ are the infinitesimal symmetries of the linear ODE, $\ddot{u}=0$.

The free particle equation admits the following eight Lie point symmetries \cite{gladwin:jpa:2011},
\begin{eqnarray} 
\Lambda_1=\frac{\partial}{\partial t},\quad            
\Lambda_2=\frac{\partial}{\partial u},\quad 
\Lambda_3=t \frac{\partial}{\partial u},\quad 
\Lambda_4=u\frac{\partial}{\partial u},\quad
\Lambda_5=u \frac{\partial}{\partial t},\nonumber\\ 
\Lambda_6=t \frac{\partial}{\partial t}
,\quad 
\Lambda_7=t^2 \frac{\partial}{\partial t}
+tu \frac{\partial}{\partial u},\quad 
\Lambda_8=tu \frac{\partial}{\partial t}
+u^2 \frac{\partial}{\partial u}.
\label {sym11}
\end{eqnarray}
Substituting the expressions for $\xi$ and $\eta$ given in (\ref{sym11}) into (\ref{arbitrary-sym}), one gets the infinitesimal symmetries $\lambda$ and $\mu$ associated with the nonlinear ODE (\ref{chap7-riccatichain1}). The respective forms are given below:
\begin{eqnarray}
&&\hspace{-2cm}\Omega_1=\frac{\partial}{\partial t},\quad\Omega_2=\frac{\sqrt{2\dot{x}+x^2}}{\dot{x}+x^2}\dot{x}\frac{\partial}{\partial x},\quad\Omega_3=\frac{\sqrt{2\dot{x}+x^2}}{\dot{x}+x^2}(t\dot{x}+x)\frac{\partial}{\partial x},\nonumber\\
&&\hspace{-2cm}\Omega_4=\frac{x(2\dot{x}+x^2)}{(\dot{x}+x^2)}\frac{\partial}{\partial x},\quad\Omega_5=\frac{x}{\sqrt{2\dot{x}+x^2}}\frac{\partial}{\partial t}-x\frac{(\dot{x}+x^2)} {\sqrt{2\dot{x}+x^2}}\frac{\partial}{\partial x},\nonumber\\
&&\hspace{-2cm}\Omega_6=t\frac{\partial}{\partial t}-x\frac{\partial}{\partial x},\quad\Omega_7=t^2\frac{\partial}{\partial t}+\frac{x^2(1-tx)} {\dot{x}+x^2}\frac{\partial}{\partial x},\nonumber\\
&&\hspace{-2cm}\Omega_8=\frac{tx}{\sqrt{2\dot{x}+x^2}}\frac{\partial}{\partial t}+\bigg(\frac{x^2\sqrt{(2\dot{x}+x^2)}} {\dot{x}+x^2}- \frac{tx(\dot{x}+x^2)} {\sqrt{2\dot{x}+x^2}}\bigg)\frac{\partial}{\partial x}.\label{ric_sca_dyn_symm}
\end{eqnarray}
One can check that the above dynamical symmetries do satisfy the symmetry invariance condition. The details are given in Appendix B.  

\subsection{Dynamical symmetries of Eq.(\ref{chap7-abelchain1})}
Now we focus our attention on Eq.(\ref{chap7-abelchain1}) that appears in the Abel chain. Since the underlying ideas are the same as given in the previous example, in the following, we summarize only the results for this equation.

From (\ref{abecond}) we can obtain 
\begin{eqnarray}
\frac{\partial x} {\partial u}=\frac{\dot{x}(3\dot{x}+x^3)^{\frac{1}{2}}} {\sqrt{2}(\dot{x}+x^3)^{\frac{3} {2}}},~~
\frac{\partial x} {\partial \dot{u}}=\frac{x^4(3\dot{x}+x^3)^{\frac{1}{2}}} {3\sqrt{2}(\dot{x}+x^3)^{\frac{5} {2}}}.
\end{eqnarray}
 Substituting these two expressions into (\ref{arbitrary-sym1}) we obtain
\begin{eqnarray}
\lambda=\xi,\qquad\mu=\eta\frac{\dot{x}(3\dot{x}+x^3)^{\frac{1}{2}}} {\sqrt{2}(\dot{x}+x^3)^{\frac{3} {2}}}+(\dot{\eta}-\dot{u}\dot{\xi})\frac{x^4(3\dot{x}+x^3)^{\frac{1}{2}}} {3\sqrt{2}(\dot{x}+x^3)^{\frac{5} {2}}}.\label{lammu}
\end{eqnarray}
Using the known Lie point symmetries (\ref{sym11}) of the free particle equation in (\ref{lammu}) we obtain the following dynamical symmetries of the nonlinear ODE (\ref{chap7-abelchain1}):
\begin{eqnarray}
&&\hspace{-0.5cm}\Omega_1=\frac{\partial}{\partial t},\quad\Omega_2=\frac{\dot{x}(3\dot{x}+x^3)^{\frac{1}{2}}} {\sqrt{2}(\dot{x}+x^3)^{\frac{3} {2}}}\frac{\partial}{\partial x},\quad\Omega_3=\frac{(3\dot{x}+x^3)^{\frac{1}{2}}} {\sqrt{2}(\dot{x}+x^3)^{\frac{5} {2}}}\bigg(t\dot{x}+\frac{x^4} {3}\bigg)\frac{\partial}{\partial x},\nonumber\\
&&\hspace{-0.5cm}\Omega_4=\frac{x}{3}\frac{(3\dot{x}+x^3)} {(\dot{x}+x^3)}\frac{\partial}{\partial x},\quad\Omega_5=\frac{\sqrt{2}x\sqrt{\dot{x}+x^3}} {\sqrt{3\dot{x}+x^3}}\frac{\partial}{\partial t}-\frac{\sqrt{2}x^4\sqrt{\dot{x}+x^3}} {3\sqrt{3\dot{x}+x^3}}\frac{\partial}{\partial x},\nonumber\\
&&\hspace{-0.5cm}\Omega_6=t\frac{\partial}{\partial t}-\frac{x^4} {3(\dot{x}+x^3)}\frac{\partial}{\partial x},\quad\Omega_7=t^2\frac{\partial}{\partial t}+\frac{2tx^4\dot{x}+3tx\dot{x}^2-tx^7+x^5} {3(\dot{x}+x^3)^2}\frac{\partial}{\partial x},\nonumber\\
&&\hspace{-0.5cm}\Omega_8=\frac{t\sqrt{2}x\sqrt{\dot{x}+x^3}} {\sqrt{3\dot{x}+x^3}}\frac{\partial}{\partial t}+\frac{\sqrt{2}x^2\big(3\dot{x}+x^2\big(x-t(\dot{x}+x^3)\big)\big)} {3\sqrt{(3\dot{x}+x^3)(\dot{x}+x^3)}}\frac{\partial}{\partial x}.
\end{eqnarray}
One can check that the above dynamical symmetries satisfy the symmetry invariance condition. 

The procedure developed above can be extended to the higher order ODEs in the Riccati and Abel chains in a straightforward manner with suitable identification of the contact transformations. The results for a third order nonlinear ODE is demonstrated in Appendix A. 

\section{Linearizing contact transformations of coupled chains}
In a recent paper, we have also reported integrable coupled Riccati and Abel chains and presented the method of deriving their general solutions \cite{gladwin:jmp:2010}. Here we derive the contact transformations, linearization and dynamical symmetries of the equations which appear in these two chains. Since the procedure is entirely algorithmic, we consider only one case each from the coupled Riccati and Abel chains, respectively, and discuss the essential ideas here. 
\subsection{General theory}
The coupled Riccati and Abel chains \cite{gladwin:jmp:2010} are, respectively, of the form
\begin{eqnarray}
\left(\frac{d}{dt}+f_1(x,y,t)\right)^mg_1(x,y,t)=0,~\left(\frac{d}{dt}+f_2(x,y,t)\right)^mg_2(x,y,t)=0\label{chap7-cmee-gen},
\end{eqnarray}
with $f_1=f_2=x+y$ for the Riccati chain and $f_1=f_2=x^2+y^2$ for the Abel chain. Equation (\ref{chap7-cmee-gen}) is connected to the linear system of ODEs,
\begin{eqnarray}
\frac{d^mu}{dt^m}=0,\qquad\frac{d^mv}{dt^m}=0,\label{chap7-linear2}
\end{eqnarray}
through the nonlocal transformation
\begin{eqnarray}
u=g_1(x,y,t)e^{\int f_1(x,y,t)dt},\qquad v=g_2(x,y,t)e^{\int f_2(x,y,t)dt},\label{chap7-nonlocal3}
\end{eqnarray}
where $f_i$'s and $g_i$'s, $i=1,2$, are functions of their arguments. From the above nonlocal transformation one can obtain the following identities,
\begin{subequations}
\begin{eqnarray}
\frac{\dot{u}}{u}=\frac{\dot{g}_1(t,x,y)+g_1(t,x,y)f_1(t,x,y)}{g_1(t,x,y)},\\
\frac{\dot{v}}{v}=\frac{\dot{g}_2(t,x,y)+g_2(t,x,y)f_2(t,x,y)}{g_2(t,x,y)}.
\end{eqnarray}
\end{subequations}
Here to explore the contact transformation, we introduce two arbitrary functions $h_1(t,x,y,\mathbf{x}^{(m-1)},\mathbf{y}^{(m-1)})$ and $h_2(t,x,y,\mathbf{x}^{(m-1)},\mathbf{y}^{(m-1)})$, where $\mathbf{x}^{(m-1)}=\{x^{(1)},x^{(2)},\ldots,x^{(m-1)}\}$ and $\mathbf{y}^{(m-1)}=\{y^{(1)},y^{(2)},\ldots,y^{(m-1)}\}$ with $x^{(n)}=\frac{d^nx}{dt^n},y^{(n)}=\frac{d^ny}{dt^n}$. 
As in the scalar case, we decompose the above system of equations into the following forms,
\begin{subequations}
\label{chap7-decompose7}
\begin{eqnarray}
u=\frac{g_1(t,x,y)}{h_1(t,x,y,\mathbf{x}^{(m-1)},\mathbf{y}^{(m-1)})},\quad \dot{u}=\frac{\dot{g}_1(t,x,y)+g_1(t,x,y)f_1(t,x,y)}{h_1(t,x,y,\mathbf{x}^{(m-1)},\mathbf{y}^{(m-1)}))},\label{chap7-decompose5}\\
v=\frac{g_2(t,x,y)}{h_2(t,x,y,\mathbf{x}^{(m-1)},\mathbf{y}^{(m-1)})},\quad \dot{v}=\frac{\dot{g}_2(t,x,y)+g_2(t,x,y)f_2(t,x,y)}{h_2(t,x,y,\mathbf{x}^{(m-1)},\mathbf{y}^{(m-1)})}.\label{chap7-decompose6}
\end{eqnarray}
\end{subequations}

Differentiating the first expression given in (\ref{chap7-decompose5}) with respect to `$t$' and substituting it into the second expression in (\ref{chap7-decompose5}) and implementing the same in (\ref{chap7-decompose6}), we find
\begin{subequations}
\label{chap7-cmee-h}
\begin{eqnarray}
\left(\frac{d}{dt}+f_1(t,x,y)\right)h_1(t,x,y,\mathbf{x}^{(m-1)},\mathbf{y}^{(m-1)})=0\label{chap7-cmee-h1},\\
\left(\frac{d}{dt}+f_2(t,x,y)\right)h_2(t,x,y,\mathbf{x}^{(m-1)},\mathbf{y}^{(m-1)})=0.\label{chap7-cmee-h2}
\end{eqnarray}
\end{subequations}
As we did in the scalar case, rewriting Eq.(\ref{chap7-cmee-h}) and comparing the resultant form with (\ref{chap7-cmee-gen}), we can show that
\begin{subequations}
\label{chap7-cmee-hg}
\begin{eqnarray}
\addtocounter{equation}{-1}
\addtocounter{equation}{1}
&&h_1=\left(\frac{d}{dt}+f_1(t,x,y)\right)^{m-1}g_1(t,x,y),\label{h1_extra}\\
&&h_2=\left(\frac{d}{dt}+f_2(t,x,y)\right)^{m-1}g_2(t,x,y)\label{h2_extra}.
\end{eqnarray}
\end{subequations}
We note here that the general solution of Eqs.(\ref{chap7-cmee-h1}) and (\ref{chap7-cmee-h2}) is a linear combination of  $h_1(t,x,y,\mathbf{x}^{(m-1)},\mathbf{y}^{(m-1)})$ and $h_2(t,x,y,\mathbf{x}^{(m-1)},\mathbf{y}^{(m-1)})$.  Therefore the denominator, say $h$, of the contact transformation (\ref{chap7-decompose7}) becomes
\begin{eqnarray}
h=Ah_1(t,x,y,\mathbf{x}^{(m-1)},\mathbf{y}^{(m-1)})+Bh_2(t,x,y,\mathbf{x}^{(m-1)},\mathbf{y}^{(m-1)}),\label{h_cou_2nd_par}
\end{eqnarray}
where $A$ and $B$ are arbitrary constants. However, as in the scalar case, to obtain a useful form of $h$, we should also include an arbitrary function (which is a function of the first integral) in the above expression for $h$.

Therefore, we choose the general solution of (\ref{chap7-cmee-hg}) to be of the form
\begin{eqnarray}
h=F(I)\left(Ah_1(t,x,y,\mathbf{x}^{(m-1)},\mathbf{y}^{(m-1)})+Bh_2(t,x,y,\mathbf{x}^{(m-1)},\mathbf{y}^{(m-1)})\right),
\end{eqnarray}
where $I$ is an integral of motion of Eq. (\ref{chap7-cmee-gen}) and $A$ and $B$ are arbitrary constants.
As a result, the contact transformation (\ref{chap7-nonlocal3}) reduces to the following form
\begin{subequations}
\begin{eqnarray}
&&\hspace{-1cm}u=\frac{F(I)g_1(x,y,t)}{A\left(\frac{d}{dt}+f_1(x,y,t)\right)g_1(x,y,t)+B\left(\frac{d}{dt}+f_2(x,y,t)\right)g_2(x,y,t)},\\
&&\hspace{-1cm}v=\frac{F(I)g_2(x,y,t)}{A\left(\frac{d}{dt}+f_1(x,y,t)\right)g_1(x,y,t)+B\left(\frac{d}{dt}+f_2(x,y,t)\right)g_2(x,y,t)},\\
&&\hspace{-1cm}\dot{u}=\frac{F(I)\left(\frac{d}{dt}+f_1(x,y,t)\right)g_1(t,x,y)}{A\left(\frac{d}{dt}+f_1(x,y,t)\right)g_1(x,y,t)+B\left(\frac{d}{dt}+f_2(x,y,t)\right)g_2(x,y,t)},\\
&&\hspace{-1cm}\dot{v}=\frac{F(I)\left(\frac{d}{dt}+f_2(x,y,t)\right)g_2(t,x,y)}{A\left(\frac{d}{dt}+f_1(x,y,t)\right)g_1(x,y,t)+B\left(\frac{d}{dt}+f_2(x,y,t)\right)g_2(x,y,t)}.
\end{eqnarray}
\end{subequations}
Thus, we find that Eq. (\ref{chap7-cmee-gen}) is connected to the uncoupled linear system of ODEs (\ref{chap7-linear2}) through the above contact transformation in addition to the nonlocal transformation (\ref{chap7-nonlocal3}).

\subsection{Coupled Riccati equation}
To illustrate the above ideas we consider a specific example of the coupled Riccati chain, namely a system of two coupled MEEs of the form
\begin{subequations}
\begin{eqnarray}
\addtocounter{equation}{-1}
\label{cmee-equation}
\addtocounter{equation}{1}
&&\ddot{x}+2(k_1x+k_2y)\dot{x}+(k_1\dot{x}+k_2\dot{y})x+
(k_1x+k_2y)^2x=0,\\
&&\ddot{y}+2(k_1x+k_2y)\dot{y}+(k_1\dot{x}+k_2\dot{y})y+
(k_1x+k_2y)^2y=0,
\end{eqnarray}
\end{subequations}
where $k_1$ and $k_2$ are two arbitrary parameters. Eq. (\ref{cmee-equation}) is obtained by fixing $g_1(t,x,y)=x$, $g_2(t,x,y)=y$ and $f_1(t,x,y)=f_2(t,x,y)=k_1x+k_2y$ in Eq. (\ref{chap7-cmee-gen}). For simplicity we take $k_1=k_2=1$ so that Eq.(\ref{cmee-equation}) becomes
\begin{subequations}
\begin{eqnarray}
\addtocounter{equation}{-1}
\label{cmee-equation1}
\addtocounter{equation}{1}
&&\ddot{x}+2(x+y)\dot{x}+(\dot{x}+\dot{y})x+
(x+y)^2x=0,\\
&&\ddot{y}+2(x+y)\dot{y}+(\dot{x}+\dot{y})y+
(x+y)^2y=0.
\end{eqnarray}
\end{subequations}
The integrability of the two MEE (\ref{cmee-equation}) was examined through the modified Prelle-Singer procedure and its general solution was reported in \cite{chandru}. Here we are interested to capture the linearizing contact transformation and dynamical symmetries of this equation. 
To deduce the linearizing contact transformation we have to find an appropriate first integral of (\ref{cmee-equation1}). As in the scalar case, this integral can be identified from the ratio of the solutions of Eqs.(\ref{chap7-cmee-hg}) and (\ref{h_cou_2nd_par}). By substituting $f_1=f_2=x+y$ and $g_1=x, g_2=y$ into Eqs.(\ref{h1_extra}) and (\ref{h2_extra}) we find that one particular solution is in the form
\begin{eqnarray}
h_1&=&\dot{x}+(x+y)x,~~h_2=\dot{y}+(x+y)y.\label{hcoura2}
\end{eqnarray}
A linear combination of the above two expressions also constitutes a solution of Eq.(\ref{chap7-cmee-h}), that is
\begin{eqnarray}
h=\dot{x}+\dot{y}+(x+y)x+(x+y)y.\label{hcombination}
\end{eqnarray}
From the ratios (i) $\frac{h_1} {h}$ and (ii) $\frac{h_2} {h}$ we can find two integrals in the forms,
\begin{subequations}
\begin{eqnarray}
\addtocounter{equation}{-1}
\label{cmee-equation1-integral}
\addtocounter{equation}{1}
&& I_1=\frac{h_1} {h_1+h_2}=\frac{\dot{x}+(x+y)x}{\dot{x}+\dot{y}+(x+y)x+(x+y)y},\\
&& I_2=\frac{h_2} {h_1+h_2}=\frac{\dot{y}+(x+y)y}{\dot{x}+\dot{y}+(x+y)x+(x+y)y}.
\end{eqnarray}
\end{subequations}
One can check that $\frac{dI_i}{dt}=0,~~i=1,2$.
\subsubsection{Contact transformations}
Unlike the scalar case, we have more flexibility in choosing $F(I)$. For example, one can consider either  $I_1/I_2$ or their combination for $I$. By substituting $g_1=x$, $g_2=y$, $f=x+y$, the associated form of $h$ becomes
\begin{equation}
h=\frac{(\dot{x}+x(x+y))(\dot{y}+y(x+y))+\dot{x}\dot{y}+xy(\dot{x}+\dot{y})} {2(\dot{x}+x(x+y)) (\dot{y}+y(x+y))}.
\end{equation}
Using the corresponding $I$, the associated contact transformation turns out to be
\begin{subequations}
\label{riccati_cou_con_tra}
\begin{eqnarray}
u&=&\frac{2x (\dot{x}+x(x+y)) (\dot{y}+y(x+y))} {(\dot{x}+x(x+y))(\dot{y}+y(x+y))+\dot{x}\dot{y}+xy(\dot{x}+\dot{y})},\\
v&=&\frac{2 y(\dot{x}+x(x+y)) (\dot{y}+y(x+y))} {(\dot{x}+x(x+y))(\dot{y}+y(x+y))+\dot{x}\dot{y}+xy(\dot{x}+\dot{y})},\\
\dot{u}&=&\frac{2 (\dot{x}+x(x+y))^2 (\dot{y}+y(x+y))} {(\dot{x}+x(x+y))(\dot{y}+y(x+y))+\dot{x}\dot{y}+xy(\dot{x}+\dot{y})},\\
\dot{v}&=&\frac{2 (\dot{x}+x(x+y)) (\dot{y}+y(x+y))^2} {(\dot{x}+x(x+y))(\dot{y}+y(x+y))+\dot{x}\dot{y}+xy(\dot{x}+\dot{y})},
\end{eqnarray}
\end{subequations}
\normalsize
or equivalently
\begin{subequations}
\label{riccati_cou_con_tra1}
\begin{eqnarray}
x&=&\frac{u} {1+\frac{1}{2}\left(\frac{u^2}{\dot{u}}+\frac{v^2}{\dot{v}} \right)},\\
y&=&\frac{v} {1+\frac{1}{2}\left(\frac{u^2}{\dot{u}}+\frac{v^2}{\dot{v}} \right)},\\
\dot{x}&=&\frac{-u(u+v)} {\bigg(1+\frac{1}{2}\left(\frac{u^2}{\dot{u}}+\frac{v^2}{\dot{v}} \right)\bigg)^2}+\frac{\dot{u}}{1+\frac{1}{2}\left(\frac{u^2}{\dot{u}}+\frac{v^2}{\dot{v}} \right)},\\
\dot{y}&=&\frac{-v(u+v)} {\bigg(1+\frac{1}{2}\left(\frac{u^2}{\dot{u}}+\frac{v^2}{\dot{v}} \right)\bigg)^2}+\frac{\dot{v}}{1+\frac{1}{2}\left(\frac{u^2}{\dot{u}}+\frac{v^2}{\dot{v}} \right)}.
\end{eqnarray}
\end{subequations}
One can straightforwardly check that the transformation (\ref{riccati_cou_con_tra}) linearizes Eq.(\ref{cmee-equation1}) as $\ddot{u}=0$ and $\ddot{v}=0$. \emph{The linearizing transformation given here is new}. Here also we can derive the solution for the associated nonlinear differential equations from the linear one. From (\ref{riccati_cou_con_tra1}) we can find the general solution of Eq.(\ref{cmee-equation1}) as 
\begin{eqnarray}
x(t)&=&\frac{2(I_1t+I_2)I_1} {1+(I_1 t+I_2)^2+(I_3 t+I_4)^2},\\
y(t)&=&\frac{2(I_3t+I_4)I_1} {1+(I_1 t+I_2)^2+(I_3 t+I_4)^2},
\end{eqnarray}
where $I_i's, i=1,2,3,4,$ are arbitrary constants.
\subsubsection{Dynamical symmetries}

Let us denote the symmetry vector field associated with the system of linear ODEs (\ref{chap7-linear2}) as
\begin{eqnarray}
\Lambda=\xi\frac{\partial}{\partial t}+\eta_1\frac{\partial}{\partial u}+\eta_2\frac{\partial}{\partial v},\label{suu1}
\end{eqnarray}
where $\xi=\xi(t,u,v)$, $\eta_1(t,u,v)$ and $\eta_2(t,u,v)$ are the Lie point symmetries of the system of free particle equations (\ref{chap7-linear2}). 
The first prolongation of the vector field (\ref{suu1}) is given by
\begin{eqnarray}
\Lambda^1=\xi\frac{\partial}{\partial t}+\eta_1\frac{\partial}{\partial u}+\eta_2\frac{\partial}{\partial v}+(\dot{\eta}_1-\dot{u}\dot{\xi})\frac{\partial}{\partial \dot{u}}+(\dot{\eta}_2-\dot{v}\dot{\xi})\frac{\partial}{\partial \dot{v}}.\label{linearprolongation}
\end{eqnarray}
Let us denote the symmetry vector field of the nonlinear system (\ref{cmee-equation1}) as
\begin{eqnarray}
\Omega=\lambda\frac{\partial}{\partial t}+\mu_1\frac{\partial}{\partial x}+\mu_2\frac{\partial}{\partial y},\label{jjklk}
\end{eqnarray}
where $\lambda=\lambda(t,x,y,\dot{x},\dot{y})$, $\mu_1(t,x,y,\dot{x},\dot{y})$ and $\mu_2(t,x,y,\dot{x},\dot{y})$ are the dynamical symmetries. The first prolongation of the vector field (\ref{jjklk}) is given by
\begin{eqnarray}
\Omega^1=\lambda\frac{\partial}{\partial t}+\mu_1\frac{\partial}{\partial x}+\mu_2\frac{\partial}{\partial y}+(\dot{\mu}_1-\dot{x}\dot{\lambda})\frac{\partial}{\partial \dot{x}}+(\dot{\mu}_2-\dot{y}\dot{\lambda})\frac{\partial}{\partial \dot{y}}.\label{nonlinearprolongation}
\end{eqnarray}




Now we construct the dynamical symmetries of (\ref{cmee-equation1}) from out of the vector field (\ref{suu1}). 
The partial derivatives that appear in the vector field (\ref{linearprolongation}) can now be expressed in terms of the variables which are present in the vector field (\ref{nonlinearprolongation}) through the following relations, 
\begin{subequations}
\label{coupled-contact2}
\begin{eqnarray}
\addtocounter{equation}{-1}
\frac{\partial}{\partial u}=\frac{\partial x}{\partial u}\frac{\partial}{\partial x}+\frac{\partial y}{\partial u}\frac{\partial}{\partial y} +\frac{\partial \dot{x}}{\partial u}\frac{\partial}{\partial \dot{x}}+\frac{\partial \dot{y}}{\partial u}\frac{\partial}{\partial \dot{y}},\nonumber\\
\frac{\partial}{\partial v}=\frac{\partial x}{\partial v}\frac{\partial}{\partial x}+\frac{\partial y}{\partial v}\frac{\partial}{\partial y}+\frac{\partial \dot{x}}{\partial v}\frac{\partial}{\partial \dot{x}}+\frac{\partial \dot{y}}{\partial v}\frac{\partial}{\partial \dot{y}}.
\end{eqnarray}
\end{subequations}
Substituting (\ref{coupled-contact2}) into the first prolongation (\ref{linearprolongation}), we find
\begin{eqnarray}
&&\Lambda^1=\xi \frac{\partial}{\partial t}+\left(\eta_1\frac{\partial x}{\partial u}+\eta_2\frac{\partial x}{\partial v}+(\dot{\eta}_1-\dot{\xi}\dot{u})\frac{\partial x}{\partial \dot{u}}+(\dot{\eta}_2-\dot{\xi}\dot{v})\frac{\partial x}{\partial \dot{v}}\right)\frac{\partial}{\partial x}
\nonumber\\
&&\hspace{1cm}+\left(\eta_1\frac{\partial y}{\partial u}+\eta_2\frac{\partial y}{\partial v}+(\dot{\eta}_1-\dot{\xi}\dot{u})\frac{\partial y}{\partial \dot{u}}+(\dot{\eta}_2-\dot{\xi}\dot{v})\frac{\partial y}{\partial \dot{v}}\right)\frac{\partial}{\partial y}
\nonumber\\
&&\hspace{1cm}+\left(\eta_1\frac{\partial \dot{x}}{\partial u}+\eta_2\frac{\partial \dot{x}}{\partial v}+(\dot{\eta}_1-\dot{\xi}\dot{u})\frac{\partial \dot{x}}{\partial \dot{u}}+(\dot{\eta}_2-\dot{\xi}\dot{v})\frac{\partial \dot{x}}{\partial \dot{v}}\right)\frac{\partial}{\partial \dot{x}}
\nonumber\\
&&\hspace{1cm}+\left(\eta_1\frac{\partial \dot{y}}{\partial u}+\eta_2\frac{\partial \dot{y}}{\partial v}+(\dot{\eta}_1-\dot{\xi}\dot{u})\frac{\partial \dot{y}}{\partial \dot{u}}+(\dot{\eta}_2-\dot{\xi}\dot{v})\frac{\partial \dot{y}}{\partial \dot{v}}\right)\frac{\partial}{\partial \dot{y}}\label{coupled-prolongation2}.
\end{eqnarray}
Comparing the prolongation vectors (\ref{nonlinearprolongation}) and (\ref{coupled-prolongation2}), we find
\begin{eqnarray}
\xi=\lambda,\quad \mu_1&=&\eta_1\frac{\partial x}{\partial u}+\eta_2\frac{\partial x}{\partial v}+(\dot{\eta}_1-\dot{\xi}\dot{u})\frac{\partial x}{\partial \dot{u}}+(\dot{\eta}_2-\dot{\xi}\dot{v})\frac{\partial x}{\partial \dot{v}},\label{dumm}\\
\mu_2&=&\eta_1\frac{\partial y}{\partial u}+\eta_2\frac{\partial y}{\partial v}+(\dot{\eta}_1-\dot{\xi}\dot{u})\frac{\partial y}{\partial \dot{u}}+(\dot{\eta}_2-\dot{\xi}\dot{v})\frac{\partial y}{\partial \dot{v}}.\label{coupsymeq2}
\end{eqnarray}

Substituting the expressions for the partial derivatives that appear in the infinitesimals $\mu_1$ and $\mu_2$ in Eqs.(\ref{dumm}) and (\ref{coupsymeq2}) and simplifying the resultant expressions, we obtain 
\begin{subequations}
\begin{eqnarray}
\addtocounter{equation}{-1}
\label{mu12}
\addtocounter{equation}{1}
\mu_1&=&a_1(2a_2a_3[\dot{x}a_3\eta_1+xy(a_3\eta_1-a_2\eta_2)]+x(x^2a_3^2\dot{\eta}_1+y^2a_2^2\dot{\eta}_2))\nonumber \\ &&\hspace{6cm}-\frac{x\dot{\xi}}{2}\bigg(\frac{x^2}{a_2}+\frac{y^2}{a_3} \bigg),\\
\mu_2&=&a_1(2a_2a_3[\dot{y}a_2\eta_2+xy(a_2\eta_2-a_3\eta_1)]+y(x^2a_3^2\dot{\eta}_1+y^2a_2^2\dot{\eta}_2))\nonumber \\ &&\hspace{6cm}-\frac{y\dot{\xi}}{2}\bigg(\frac{x^2}{a_2}+\frac{y^2}{a_3} \bigg),
\end{eqnarray}
\end{subequations}
where $a_i,~i=1,2,3$, are defined by
\begin{eqnarray}
&&a_1=(\dot{x}+x(x+y))(\dot{y}+y(x+y))+\dot{x}\dot{y}+xy(\dot{x}+\dot{y}),\nonumber \label{sim1}\\
&&a_2=\dot{x}+x(x+y),~~ a_3=\dot{y}+y(x+y).\label{sim3}
\end{eqnarray}

Since we know the infinitesimal symmetries of the linearized system of ODEs, $\ddot{u}=0$ and $\ddot{v}=0$, they can be readily substituted into Eq.(\ref{mu12}) to provide the infinitesimal symmetries of the nonlinear ODE (\ref{cmee-equation1}). These infinitesimals not only contain the variables $t$, $x$ and $y$ but also $\dot{x}$ and $\dot{y}$, as we see below.


The symmetry vector fields of the system of linear ODEs are given by \cite{gladwin:jpa:2011}
\begin{eqnarray}
&&\Lambda_1=\frac{\partial}{\partial t},\quad\Lambda_2=\frac{\partial}{\partial u},\quad\Lambda_3=\frac{\partial}{\partial v}\quad\Lambda_4=t\frac{\partial}{\partial t},\quad\Lambda_5=t\frac{\partial}{\partial u},\quad\Lambda_6=t\frac{\partial}{\partial v},\nonumber\\
&&\Lambda_7=u\frac{\partial}{\partial t},\quad\Lambda_8=v\frac{\partial}{\partial t},\quad\Lambda_9=u\frac{\partial}{\partial v},\quad\Lambda_{10}=v\frac{\partial}{\partial u},\quad\Lambda_{11}=u\frac{\partial}{\partial u}+v\frac{\partial}{\partial v},\nonumber\\
&&\Lambda_{12}=u\frac{\partial}{\partial u}-v\frac{\partial}{\partial v},\quad\Lambda_{13}=t^2\frac{\partial}{\partial t}+ut\frac{\partial}{\partial u}+vt\frac{\partial}{\partial v},\nonumber\\
&&\Lambda_{14}=ut\frac{\partial}{\partial t}+u^2\frac{\partial}{\partial u}+uv\frac{\partial}{\partial v},\quad\Lambda_{15}=vt\frac{\partial}{\partial t}+uv\frac{\partial}{\partial u}+v^2\frac{\partial}{\partial v}.\label{lie-point}
\end{eqnarray}
For each one of the symmetries given in (\ref{lie-point}) we can identify a contact transformation. For the sake of completeness we demonstrate the method of identifying contact symmetries by considering only one vector field, for example $\Lambda_9$. One can repeat the procedure and obtain the contact symmetries for the rest of the vector fields. In the case of $\Lambda_9$, we have $\eta_1=0,~~\eta_2=u,~~\xi=0$. Substituting these forms in (\ref{dumm}) and (\ref{coupsymeq2}) and simplifying the resultant expressions, we find the contact symmetry $\Omega_9$ which is given in Eq.(\ref{cont_couip_ex}) below. Similarly, substituting all the other symmetries given in (\ref{lie-point}) into Eqs.(\ref{dumm}) and (\ref{coupsymeq2}) and simplifying them appropriately, we get the following dynamical symmetries of Eq.(\ref{cmee-equation1}):
\begin{eqnarray}
&&\Omega_1=\frac{\partial}{\partial t},\quad\Omega_2=\frac{a_1}{2a_2^2a_3}\bigg((\dot{x}+xy)\frac{\partial}{\partial x}-xy\frac{\partial}{\partial y}\bigg),\nonumber\\
&&\Omega_3=\frac{a_1}{2a_2a_3^2}\bigg(-xy\frac{\partial}{\partial x}+(\dot{y}+xy)\frac{\partial}{\partial y}\bigg),\nonumber\\
&&\Omega_4=t\frac{\partial}{\partial t}-\bigg(\frac{a_2 y^2+a_3x^2}{2a_2a_3}\bigg)\bigg(x\frac{\partial}{\partial x}+y\frac{\partial}{\partial y}\bigg),\nonumber\\
&&\Omega_5=\frac{a_1}{4 a_2^3a_3}\bigg((2ta_2(\dot{x}+xy)+x^3)\frac{\partial}{\partial x}-xy(2ta_2-x)\frac{\partial}{\partial y}\bigg),\nonumber\\
&&\Omega_6=\frac{a_1}{4 a_2a_3^3}\bigg(xy(y-2ta_3)\frac{\partial}{\partial x}+(2ta_3(\dot{y}+xy)+y^3)\frac{\partial}{\partial y}\bigg),\quad\nonumber\\
&&\Omega_7=\frac{a_2}{a_1}\bigg(2xa_3\frac{\partial}{\partial t}-(a_3x^2+a_2y^2)\bigg)\bigg(x\frac{\partial}{\partial x}+y\frac{\partial}{\partial y}\bigg),\nonumber\\
&&\Omega_8=\frac{a_3}{a_1}\bigg(2ya_2\frac{\partial}{\partial t}-(a_3x^2+a_2y^2)\bigg)\bigg(x\frac{\partial}{\partial x}+y\frac{\partial}{\partial y}\bigg),\nonumber\\
&&\Omega_9=-\frac{xy(xy(x+y)-\dot{x}y+2x\dot{y})}{2a_3^2}\frac{\partial}{\partial x}+\bigg(\frac{x(\dot{y}+xy)}{a_3}+\frac{a_2y^3}{2a_3^2}\bigg)\frac{\partial}{\partial y},\nonumber\\
&&\Omega_{10}=\bigg(\frac{y(\dot{x}+xy)}{a_2}+\frac{a_3x^3}{2a_2^2}\bigg)\frac{\partial}{\partial x}-\frac{xy(xy(x+y)-x\dot{y}+2y\dot{x})}{2a_2^2}\frac{\partial}{\partial y},\nonumber\\
&&\Omega_{11}=\frac{a_1}{2a_2a_3}\bigg(x\frac{\partial}{\partial x}+y\frac{\partial}{\partial y}\bigg),\nonumber\\
&&\Omega_{12}=\bigg(\frac{xa_3(\dot{x}+xy)+x^2ya_2}{a_2a_3}+\frac{x(x^2a_3^2-y^2a_2^2)}{2a_2a_3^2}\bigg)\frac{\partial}{\partial x}\nonumber \\
&& +\bigg(\frac{y(x^2a_3^2-y^2a_2^2)}{2a_2a_3^2}-\frac{xa_2(\dot{y}+xy)+x^2ya_3}{a_2a_3}+\bigg)\frac{\partial}{\partial y},\nonumber\\
&&\Omega_{13}=t^2\frac{\partial} {\partial t}+\bigg( \frac{t} {a_2a_3}[xa_3(\dot{x}+xy)-xy^2a_2]+\frac{x^3} {2a_2^2}(-a_2t+x)-\frac{xy^2}{2a_3^2}(a_3t-y)\bigg) \frac{\partial}{\partial x} \nonumber \\
&& + \bigg( \frac{t} {a_2a_3}[ya_3(\dot{y}+xy)-yx^2a_3]+\frac{x^2y} {2a_2^2}(-a_2t+x)+\frac{y^3}{2a_3^2}(-a_3t+y)\bigg) \frac{\partial}{\partial y},\nonumber 
\end{eqnarray}
\begin{eqnarray}
\Omega_{14}&=&\frac{2t xa_2a_3} {a_1}\frac{\partial} {\partial t}+\bigg( \frac{2x^2} {a_1}[a_3(\dot{x}+xy)-y^2a_2]+\frac{x^3a_3} {a_1}(-a_2t+x)-\frac{xy^2a_2^2}{a_1a_3}(a_3t-y)\bigg) \frac{\partial}{\partial x}\nonumber \\
&& +\bigg( \frac{2xy} {a_1}[a_2(\dot{y}+xy)-x^2a_3]+\frac{x^2ya_3} {a_1}(-a_2t+x)+\frac{y^3a_2^2}{a_1a_3}(-a_3t+y)\bigg) \frac{\partial}{\partial y},\nonumber\\
\Omega_{15}&=&\frac{2t ya_2a_3} {a_1}\frac{\partial} {\partial t}\nonumber +\bigg( \frac{2xy} {a_1}[a_3(\dot{x}+xy)-y^2a_2]+\frac{x^3a_3^2} {a_1a_2}(-a_2t+x)-\frac{xy^2a_2}{a_1}(a_3t-y)\bigg)\frac{\partial}{\partial x}\nonumber \\
&& +\bigg( \frac{2y^2} {a_1}[a_2(\dot{y}+xy)-x^2a_3]+\frac{x^2ya_3^2} {a_1a_2}(-a_2t+x)+\frac{y^3a_2}{a_1}(-a_3t+y)\bigg)\frac{\partial}{\partial y},\label{cont_couip_ex}
\end{eqnarray}
where $a_1,~a_2$ and $a_3$ are given in Eq.(\ref{sim3}). One can unambiguously check the symmetries reported above do satisfy the invariance condition.

\subsection{Coupled Abel equation}
The theory developed in the previous section is not only applicable to the coupled Riccati chain but also to the coupled Abel chain. To demonstrate this feature we consider the two coupled generalization of Abel equations and derive the linearizing contact transformation and dynamical symmetries of it. The two coupled generalization of Abel's equation is given by
\begin{eqnarray}
\ddot{x}+2(k_1x^2+k_2y^2)\dot{x}+((k_1x^2+k_2y^2)^2+2 k_1 x \dot{x}+2 k_2 y \dot{y})x=0,\nonumber \\
\ddot{y}+2(k_1x^2+k_2y^2)\dot{y}+((k_1x^2+k_2y^2)^2+2 k_1 x \dot{x}+2 k_2 y \dot{y})y=0.
\end{eqnarray} 
Again for simplicity, hereafter we consider $k_1, k_2=1$. Then the above equation of motion becomes 
\begin{eqnarray}
\ddot{x}+2(x^2+y^2)\dot{x}+((x^2+y^2)^2+2 x \dot{x}+2 y \dot{y})x=0,\nonumber \\
\ddot{y}+2(x^2+y^2)\dot{y}+((x^2+y^2)^2+2 x \dot{x}+2 y \dot{y})y=0.\label{abe_coe_eq}
\end{eqnarray} 
To deduce the linearizing contact transformation of (\ref{abe_coe_eq}) we have to find an appropriate first integral. For this purpose we proceed as follows. By substituting $f_1=f_2=x^2+y^2$ and $g_1=x, g_2=y$ into Eq.(\ref{chap7-cmee-hg}) we get one particular solution which is of the form
\begin{eqnarray}
h_1=\dot{x}+(x^2+y^2)x,~~h_2=\dot{y}+(x^2+y^2)y.\label{hcoura2}
\end{eqnarray}
The combination of the above two forms is also a solution of Eq.(\ref{chap7-cmee-hg}), that is
\begin{eqnarray}
h=\dot{x}+\dot{y}+(x^2+y^2)x+(x^2+y^2)y.\label{hcombination}
\end{eqnarray}
From the ratios (i) $\frac{h_1} {h}$ and (ii) $\frac{h_2} {h}$ we can find the necessary integrals as
\begin{eqnarray}
I_1=\frac{\dot{x}+(x^2+y^2)x}{\dot{x}+\dot{y}+(x^2+y^2)x+(x^2+y^2)y},\nonumber \\
I_2=\frac{\dot{y}+(x^2+y^2)y}{\dot{x}+\dot{y}+(x^2+y^2)x+(x^2+y^2)y},
\end{eqnarray}
which are nothing but the ratio of the particular solutions of (\ref{chap7-cmee-h}).
One can check that $\frac{dI_i}{dt}=0,~~i=1,2$. Once suitable integrals are derived the necessary linearizing contact transformations can be identified from which one can derive the dynamical symmetries of the coupled nonlinear ODE (\ref{abe_coe_eq}). Since the algorithm is essentially the same as in the case of two coupled Riccati equations, in the following, we summarize only the results.
\subsubsection{Contact transformation}
Extending the analysis given in the previous subsection, we find the following expression for $h$, that is
\begin{equation}
h=\frac{[x(y(x^2+y^2)^2+\dot{y}(x^2+3 y^2))+\dot{x}(3\dot{y}+3x^2y+y^3)]^{\frac{1}{2}}} {[3 (\dot{x}+x(x^2+y^2)) (\dot{y}+y(x^2+y^2))]^{\frac{1}{2}}}
\end{equation}
and the associated linearizing contact transformation turns out to be
\begin{subequations}
\label{trans_abel_cou}
\begin{eqnarray}
u&=&\frac{x[3 (\dot{x}+x(x^2+y^2)) (\dot{y}+y(x^2+y^2))]^{\frac{1}{2}}} {[x(y(x^2+y^2)^2+\dot{y}(x^2+3 y^2))+\dot{x}(3\dot{y}+3x^2y+y^3)]^{\frac{1}{2}}}, \\
v&=&\frac{y[3 (\dot{x}+x(x^2+y^2)) (\dot{y}+y(x^2+y^2))]^{\frac{1}{2}}} {[x(y(x^2+y^2)^2+\dot{y}(x^2+3 y^2))+\dot{x}(3\dot{y}+3x^2y+y^3)]^{\frac{1}{2}}},\\
\dot{u}&=&\frac{x(\dot{x}+x(x^2+y^2))[3 (\dot{x}+x(x^2+y^2)) (\dot{y}+y(x^2+y^2))]^{\frac{1}{2}}} {[x(y(x^2+y^2)^2+\dot{y}(x^2+3 y^2))+\dot{x}(3\dot{y}+3x^2y+y^3)]^{\frac{1}{2}}},\\
\dot{v}&=&\frac{y(\dot{y}+y(x^2+y^2))[3 (\dot{x}+x(x^2+y^2)) (\dot{y}+y(x^2+y^2))]^{\frac{1}{2}}} {[x(y(x^2+y^2)^2+\dot{y}(x^2+3 y^2))+\dot{x}(3\dot{y}+3x^2y+y^3)]^{\frac{1}{2}}},
\end{eqnarray}
\end{subequations}
or equivalently
\begin{subequations}
\begin{eqnarray}
\label{trans_abel_cou_xy}
x&=&\frac{u} {(1+\frac{2} {3}(\frac{u^3} {\dot{u}}+\frac{v^3} {\dot{v}}))^{\frac{1} {2}}},\\
y&=&\frac{v} {(1+\frac{2} {3}(\frac{u^3} {\dot{u}}+\frac{v^3} {\dot{v}}))^{\frac{1} {2}}},\\
\dot{x}&=&-\frac{u(u^2+v^2)} {(1+\frac{2} {3}(\frac{u^3} {\dot{u}}+\frac{v^3} {\dot{v}}))^{\frac{3} {2}}}+\frac{\dot{u}} {(1+\frac{2} {3}(\frac{u^3} {\dot{u}}+\frac{v^3} {\dot{v}}))^{\frac{1} {2}}},\\
\dot{y}&=&-\frac{v(u^2+v^2)} {(1+\frac{2} {3}(\frac{u^3} {\dot{u}}+\frac{v^3} {\dot{v}}))^{\frac{3} {2}}}+\frac{\dot{v}} {(1+\frac{2} {3}(\frac{u^3} {\dot{u}}+\frac{v^3} {\dot{v}}))^{\frac{1} {2}}}.
\end{eqnarray}
\end{subequations}
One may check that the contact transformation (\ref{trans_abel_cou}) linearizes the nonlinear ODEs (\ref{abe_coe_eq}) to free particle equations $\ddot{u}=0$ and $\ddot{v}=0$. \emph {The linearizing transformation given here is also new}. Here also we can derive the solution for the nonlinear ODEs (\ref{abe_coe_eq}) from the linear one. We find that the general solution of Eq. (\ref{abe_coe_eq}) is of the form
\begin{eqnarray}
x(t)&=&\frac{I_1 t+I_2} {\bigg(1+\frac{2} {3}\bigg(\frac{(I_1t+I_2)^3} {I_1}+\frac{(I_3t+I_4)^3} {I_3}\bigg)\bigg)^{\frac{1} {2}}},\\
y(t)&=&\frac{I_3 t+I_4} {\bigg(1+\frac{2} {3}\bigg(\frac{(I_1t+I_2)^3} {I_1}+\frac{(I_3t+I_4)^3} {I_3}\bigg)\bigg)^{\frac{1} {2}}},
\end{eqnarray}
where $I_i's,i=1,2,3,4,$ are arbitrary constants.
\subsubsection{Dynamical symmetries}
We find that the infinitesimal symmetries ($\lambda,~\mu_1$,~$\mu_2$) of the nonlinear ODEs (\ref{abe_coe_eq}) can be related to the Lie point symmetries of the linear ODEs ($\xi, \eta_1$ and $\eta_2$) through the relation
\begin{eqnarray}
\xi=\lambda,~~\mu_1&=& A_1\bigg[\frac{\eta_1(\dot{x}+xy^2)} {\sqrt{3}(A_{2})^3 A_{3}} -\frac{\eta_2 xy^2}{\sqrt{3} A_{2}(A_{3})^{3}}+\frac{\dot{\eta_1}x^4}{3\sqrt{3}(A_{2})^{5} A_{3}}\nonumber \\ &&+\frac{\dot{\eta_2}y^3x}{3\sqrt{3} A_{2}(A_{3})^{5}}\bigg]-\frac{\dot{\xi}x}{3}\bigg(\frac{y^3}{(A_3)^2}+\frac{x^3}{(A_2)^2}\bigg),\nonumber \\
\mu_2&=& A_1\bigg[-\frac{\eta_1 x^2y} {\sqrt{3}(A_{2})^{3}A_{3}} +\frac{\eta_2(\dot{y}+x^2y)}{\sqrt{3}A_{2}(A_{3})^{3}}+\frac{\dot{\eta_1}x^3y}{3\sqrt{3}(A_{2})^{5}A_{3}}\nonumber \\ &&+\frac{\dot{\eta_2}y^4}{3\sqrt{3}A_{2}(A_{3})^{5}}\bigg]-\frac{\dot{\xi}y}{3}\bigg(\frac{y^3}{(A_3)^2}+\frac{x^3}{(A_2)^2}\bigg),\label{muabelcou}
\end{eqnarray}
where $A_i's, i=1,2,3$ are given by
\begin{eqnarray}
A_1&=&\sqrt{x(y(x^2+y^2)^2+\dot{y}(x^2+3 y^2))+\dot{x}(3\dot{y}+3x^2y+y^3)},\nonumber \\
A_2&=&\sqrt{x(x^2+y^2)+\dot{x}},~~A_3=\sqrt{y(x^2+y^2)+\dot{y}}.
\end{eqnarray}
Now recalling the infinitesimal symmetries of the ODE $\ddot{u}=0$ and $\ddot{v}=0$ (vide Eq.(\ref{lie-point})) and substituting them in (\ref{muabelcou}) and simplifying the resultant expressions we arrive at the following form of dynamical symmetries for the nonlinear ODEs (\ref{abe_coe_eq}):  
\begin{eqnarray}
\Omega_1&=&\frac{\partial} {\partial t},~~\Omega_2=\frac{A_1 (\dot{x}+x y^2)}{\sqrt{3}(A_2)^3A_3}\frac{\partial}{\partial x}-\frac{x^2 yA_1} {\sqrt{3}(A_2)^3A_3} \frac{\partial}{\partial y},\nonumber \\
\Omega_3&=&\frac{-x y^2A_1 }{\sqrt{3}A_2 (A_3)^3}\frac{\partial}{\partial x}+\frac{(\dot{y}+x^2 y)A_1} {\sqrt{3} A_2 (A_3)^3 } \frac{\partial}{\partial y},\nonumber \\
\Omega_4&=&t \frac{\partial} {\partial t}-\bigg(\frac{xy^3}{3 (A_3)^2}+\frac{x^4}{3(A_2)^2}\bigg)\frac{\partial} {\partial x}-\bigg(\frac{x^3y}{3 (A_2)^2}+\frac{y^4}{3(A_3)^2}\bigg)\frac{\partial} {\partial y},\nonumber\\
\Omega_5&=&\bigg(\frac{A_1 (x^4+3t(\dot{x}+xy^2)A_2)} {3\sqrt{3}A_3 (A_2)^5}\bigg)\frac{\partial} {\partial x}+\bigg(\frac{A_1(x^3 y+3tx^2 y A_2)}{3\sqrt{3} (A_2)^5A_3}\bigg)\frac{\partial} {\partial y},\nonumber\\
\Omega_6&=&\bigg(\frac{A_1 (xy^3-3txy^2A_3)} {3\sqrt{3}(A_2)(A_3)^5}\bigg)\frac{\partial} {\partial x}+\bigg(\frac{A_1(y^4+3t(\dot{y}+x^2 y)A_2)}{3\sqrt{3} A_2 (A_3)^5}\bigg)\frac{\partial} {\partial y},\nonumber \\
\Omega_7&=&\frac{\sqrt{3}xA_2A_3} {A_1}\frac{\partial} {\partial t}-\frac{x^4 A_2 (A_3)^2+x y^3 (A_2)^3} {\sqrt{3}A_2 A_3}\frac{\partial} {\partial x} -\frac{(x^3yA_2 (A_3)^2+y^4 (A_2)^3)} {\sqrt{3}A_2 A_3}\frac{\partial} {\partial y},\nonumber
\end{eqnarray}
\begin{eqnarray}
\Omega_8&=&\frac{\sqrt{3}yA_2A_3} {A_1}\frac{\partial} {\partial t}-\frac{x^4(A_3)^3+x y^3 (A_2)^2 A_3} {\sqrt{3}A_2 A_3}\frac{\partial} {\partial x} -\frac{(x^3y (A_3)^3+y^4(A_2)^2 A_3)} {\sqrt{3}A_2 A_3}\frac{\partial} {\partial y},\nonumber\\
\Omega_9&=&\bigg(\frac{-x^2y^2} {(A_3)^2}+\frac{x y^3 A_2}{3 (A_3)^4}\bigg)\frac{\partial} {\partial x}+\bigg(\frac{x(\dot{y}+x^2 y)} {(A_3)^2}+\frac{y^4 (A_2)^2} {\sqrt{3} (A_3)^4 }\bigg)\frac{\partial} {\partial y},\nonumber\\
\Omega_{10}&=&\bigg(\frac{y(\dot{x}+x y^2)} {(A_2)^2}+\frac{x^4 (A_3)^2}{3 (A_2)^4}\bigg)\frac{\partial} {\partial x}-\bigg(\frac{x^2y^2} {(A_2)^2}-\frac{x^3y (A_3)^2} {\sqrt{3} (A_2)^4 }\bigg)\frac{\partial} {\partial y},\nonumber\\
\Omega_{11}&=&\bigg(\frac{3x(\dot{x}+x y^2)+x^4} {3 (A_2)^2}-\frac{2} {3}\frac{x y^3}{ (A_3)^2} \bigg)\frac{\partial} {\partial x}+\bigg(\frac{3y(\dot{y}+x^2 y)+y^4} {3 (A_3)^2}-\frac{2} {3}\frac{x^3y}{(A_2)^2} \bigg) \frac{\partial} {\partial y},\nonumber\\
\Omega_{12}&=&\bigg(\frac{3x(\dot{x}+x y^2)+x^4} {3 (A_2)^2}+\frac{2} {3}\frac{x y^3}{ (A_3)^2} \bigg)\frac{\partial} {\partial x}-\bigg(\frac{3y(\dot{y}+x^2 y)+y^4} {3 (A_3)^2}+\frac{2} {3}\frac{x^3y}{(A_2)^2} \bigg) \frac{\partial} {\partial y},\nonumber\\
\Omega_{13}&=&t^2\frac{\partial} {\partial t}+\frac{x}{3}\bigg[\frac{x^4}{(A_2)^4}+\frac{y^4}{(A_3)^4}+t\bigg(\frac{3((\dot{x}+xy^2)-y^3)}{A_2}-\bigg(\frac{x^3}{(A_2)^2}+\frac{y^3}{(A_3)^2}\bigg)\bigg)\bigg]\frac{\partial} {\partial x}\nonumber \\ && +\frac{y}{3}\bigg[\frac{x^4}{(A_2)^4}+\frac{y^4}{(A_3)^4}+t\bigg(\frac{3((\dot{y}+x^2y)-x^3)}{A_2}-\bigg(\frac{x^3}{(A_2)^2}+\frac{y^3}{(A_3)^2}\bigg)\bigg)\bigg]\frac{\partial} {\partial y},\nonumber\\
\Omega_{14}&=&\frac{\sqrt{3}xt A_2 A_3}{A_1}\frac{\partial} {\partial t}+\frac{1} {A_1}\bigg(\frac{\sqrt{3}x^2(\dot{x}+xy^2)A_3)}{A_2}-\frac{\sqrt{3}x^2y^3A_2}{A_3 }+\frac{x^5 A_3}{\sqrt{3} A_2 }-\frac{tx^4 A_2 A_3}{\sqrt{3}}\nonumber \\ && +\frac{y^4x (A_2)^3}{\sqrt{3}(A_3)^3}-\frac{tx y^3 (A_2)^3}{\sqrt{3}A_3} \bigg)\frac{\partial} {\partial x} +\frac{1} {A_1}\bigg(\frac{\sqrt{3}xy(\dot{y}+x^2y)A_2}{A_3}-\frac{\sqrt{3}x^4yA_3}{A_2}+\frac{x^4y A_3}{\sqrt{3} A_2 }\nonumber \\ && -\frac{tx^3 y A_2 A_3}{\sqrt{3}}+\frac{y^5 (A_2)^3}{\sqrt{3}(A_3)^3}-\frac{ty^4 (A_2)^3}{\sqrt{3}A_3} \bigg)\frac{\partial} {\partial y},\nonumber\\
\Omega_{15}&=&\frac{\sqrt{3}yt A_2 A_3}{A_1}\frac{\partial} {\partial t}+\frac{1} {A_1}\bigg(\frac{\sqrt{3}xy(\dot{x}+xy^2)A_3)}{A_2}-\frac{\sqrt{3}xy^4A_2}{A_3 }+\frac{xy^4 A_2}{\sqrt{3} A_3 }-\frac{tx y^3 A_2 A_3}{\sqrt{3}}\nonumber \\ && +\frac{x^4y (A_3)^3}{\sqrt{3}(A_2)^3}-\frac{tx^4 (A_3)^3}{\sqrt{3}A_2}\bigg)\frac{\partial} {\partial x} +\frac{1} {A_1}\bigg(\frac{\sqrt{3}y^2(\dot{y}+x^2y)A_2}{A_3}-\frac{\sqrt{3}x^3y^2A_3}{A_2}+\frac{y^5 A_2}{\sqrt{3} A_3 }\nonumber \\ && -\frac{ty^4 A_2 A_3}{\sqrt{3}}+\frac{x^4y (A_3)^3}{\sqrt{3}(A_2)^3}-\frac{tx^3y (A_3)^3}{\sqrt{3}A_2} \bigg)\frac{\partial} {\partial y}.
\end{eqnarray}
In a similar fashion one can extend the theory to all the coupled equations in the Riccati and Abel chains. 
\section{Conclusion}
In this work, we have developed an algorithm to determine linearizing contact transformations and to obtain dynamical symmetries of any equation in the Riccati and Abel chains. The contact transformations and dynamical symmetries reported in this paper are new to the literature. We have also presented a method of finding the general solution of the associated nonlinear ODE. Further, the applicability of this method to higher order as well as coupled nonlinear ODEs is also illustrated explicitly. Exploring the general solution of the coupled nonlinear ODEs is often cumbersome and the algorithm proposed in this paper helps to derive the general solution in a simple and straightforward manner. The results which are presented above not only add further knowledge on these two chains of equations but also reveal the geometrical richness that these two chains possess. The extension of this method to other classes of equations is under investigation. 

\section*{Acknowledgements}
RMS acknowledges the University Grants Commission (UGC-RFSMS), Government of India, for providing a Research Fellowship and the work of MS forms part of a research project sponsored by the Department of Science and Technology, Government of India. The work of VKC is supported by the SERB-DST Fast Track scheme for young scientists under Grant No. YSS/2014/000175. The work of ML is supported by a DAE Raja Ramanna Fellowship. 
\appendix
\begin{appendix}
\section{Contact transformations and dynamical symmetries of a third order nonlinear ODE}
To confirm that the method developed in Sec.2 is also applicable to higher order nonlinear ODEs, we consider the third order nonlinear ODE in the Riccati chain, namely
\begin{equation}
\dddot{x}+4x\ddot{x}+3\dot{x}^2+6x^2\dot{x}+x^4=0,\label{thir_ric}
\end{equation}
and derive the contact transformation and dynamical symmetries of it. As mentioned in Sec.2, the function $h(t,x,\dot{x})$ can be fixed by solving Eqs.(\ref{hsol111}) and (\ref{chap7-eq-h211}), respectively, with $f=g=x$ and $m=3$. We find $h_1=\ddot{x}+3 x \dot{x}+x^3$ and $h_2=\frac{(\ddot{x}+3 x \dot{x}+x^3)^2} {\dot{x}+x^2-t(\ddot{x}+3 x \dot{x}+x^3)}$. With these expressions the integral turns out to be
\begin{equation}
I=\frac{h_1} {h_2}=\frac{\dot{x}+x^2} {\ddot{x}+3 x \dot{x}+x^3}-t.
\end{equation} 
With this form of integral $I$, we can fix the contact transformation as
\begin{eqnarray}
u&=&\frac{1} {\ddot{x}+3 x \dot{x}+x^3}-\frac{t^3} {6},\\
\dot{u}&=&\frac{x} {\ddot{x}+3 x \dot{x}+x^3}-\frac{t^2} {2},\\
\ddot{u}&=&\frac{\dot{x}+x^2} {\ddot{x}+3 x \dot{x}+x^3}-t.
\end{eqnarray}
Using the above transformation and following the procedure given in Sec.4, we can derive the contact symmetries of Eq.(\ref{thir_ric}). The resultant expressions read
\begin{eqnarray}
\Omega_1&=&\frac{\partial} {\partial t},\quad \Omega_2=-x(\ddot{x}+3 x \dot{x}+x^3)\frac{\partial} {\partial x},\nonumber\\
\Omega_3&=&(2t-t^2x)(\ddot{x}+3 x \dot{x}+x^3)\frac{\partial} {\partial x},\nonumber \\
\Omega_4&=&t\frac{\partial} {\partial t}-(x+\frac{t^2}{2}(-3+tx)(\ddot{x}+3 x \dot{x}+x^3))\frac{\partial} {\partial x},\nonumber \\
\Omega_5&=&(1-tx)(\ddot{x}+3 x \dot{x}+x^3)\frac{\partial} {\partial x},\nonumber \\
\Omega_6&=&\frac{t^2} {6}(-3+tx)(\ddot{x}+3 x \dot{x}+x^3)\frac{\partial} {\partial x},\nonumber \\
\Omega_7&=&\frac{t^2} {2}\frac{\partial} {\partial t}+((-1+tx)+\frac{t^3} {12}(4-tx)(\ddot{x}+3 x \dot{x}+x^3))\frac{\partial} {\partial x}.
\end{eqnarray}
One can check that the above dynamical symmetries indeed satisfy the symmetry invariance condition by following the details given in the following Appendix B. 

\section{Verifying the invariance condition}
In this section, we prove that the dynamical symmetries obtained in Sec.4 satisfy the invariance condition. To begin with, as an example, we consider the dynamical symmetry vector field $\Omega_4$ in (\ref{ric_sca_dyn_symm}) of Eq.(\ref{chap7-riccatichain1}):
\begin{eqnarray}
\Omega_4=\frac{x(2\dot{x}+x^2)} {\dot{x}+x^2}\frac{\partial}{\partial x}.\label{sym_ver}
\end{eqnarray}
The symmetry invariance condition is given as
\begin{eqnarray}
\bigg(\lambda\frac{\partial \phi}{\partial t}+\mu\frac{\partial \phi}{\partial x}+\mu^{(1)}\frac{\partial \phi}{\partial \dot{x}}+\mu^{(2)}\frac{\partial \phi}{\partial \ddot{x}}\bigg)|_{(\ddot{x}-\phi(x,\dot{x})=0)}=0,\label{inv_appen_1}
\end{eqnarray}
where $\phi(x,\dot{x})=-(3x\dot{x}+x^3)$, $\mu^{(1)}=\dot{\mu}-\dot{x}\dot{\lambda}$ and $\mu^{(2)}=\frac{d} {dt}{\mu^{(1)}}-\ddot{x}\dot{\lambda}$. For the symmetry vector field (\ref{sym_ver}), we have
\begin{eqnarray}
\lambda=0,\quad \mu=\frac{x(2\dot{x}+x^2)} {\dot{x}+x^2}.\label{fhjdj}
\end{eqnarray}
Therefore we find that $\mu^{(1)}=\frac{d\mu} {dt}$ and $\mu^{(2)}=\frac{d\mu^{(1)}} {dt}$. Substituting these expressions into the symmetry invariance condition (\ref{inv_appen_1}), we find
\begin{eqnarray}
-\mu\phi_x-\mu^{(1)}\phi_{\dot{x}}+\mu^{(2)}=0.\label{hlkgj}
\end{eqnarray}
Substituting the expressions $\mu$ and its derivatives into Eq. (\ref{hlkgj}), we find that the invariance condition (\ref{hlkgj}) is identically satisfied. We have also similarly verified that all the other remaining dynamical symmetries of Eq.(\ref{chap7-riccatichain1}) satisfy the invariance condition (\ref{inv_appen_1}).

\end{appendix}


\begin{thebibliography}{10}
 
\bibitem{olver}
Olver PJ. Equivalence, invariants and symmetry. Cambridge: Cambridge University Press; 1995.

\bibitem{steeb:book:93:01}
Steeb WH. Invertible point transformations and nonlinear differential equations. London: World Scientific; 1993.

\bibitem{19}
Lie S. Vorlesungen $\ddot{u}$ber Differentialgleichungen mit bekannten infinitesimalen Transformationen. Teubner: Leipzig; 1912. 

\bibitem{duarte:94:01}
Duarte LGS, Moreira IC, Santos FC. Linearization under nonpoint transformations. J. Phys. A: Math. Gen. 1994;27:L739-L743. 

\bibitem{berkovich}
Berkovich LM, Orlova IS. The exact linearization of some classes of ordinary differential equations for order $n>2$. Proc. Inst. Math. NAS Ukraine. 2000;30:90-98.

\bibitem{euler:2003}
Euler N, Wolf T, Leach PGL, Euler M. Linearisable third-order ordinary differential equations and generalised Sundman transformations : The case $\dddot{x}=0$. Acta Appl. Math. 2003;76:89-115.

\bibitem{chandrasekar:06:01:jpa}
Chandrasekar VK, Senthilvelan M, Kundu A, M. Lakshmanan. A nonlocal connection between certain linear and nonlinear differential equations/oscillators. J. Phys. A: Math. Gen. 2006;39:9743-9754.

\bibitem{gladwin:jpa:2011}
Gladwin Pradeep R, Chandrasekar VK, Senthilvelan M, Lakshmanan M. Nonlocal symmetries of a
class of scalar and coupled nonlinear ordinary differential equations of any order. J. Phys. A: Math. Theor. 2011;44:445201.

\bibitem{gladwin:jmp:2010}
Gladwin Pradeep R, Chandrasekar VK, Senthilvelan M, Lakshmanan M.  A nonlocal connection between certain linear and nonlinear ordinary differential equations: Extension to coupled equations. J. Math. Phys. 2010;51:103513-18.


\bibitem{gladwin:jpa:2006}
Chandrasekar VK, Senthilvelan M, Lakshmanan M. A unification in the theory of linearization of second-order nonlinear ordinary differential equations. J. Phys. A: Math. Gen. 2006;39:L69.



\bibitem{mahomed:1989a}
Mahomed FM, Leach PGL. Lie algebras associated with scalar second-order ordinary differential equations. J. Math. Phys. 1989;30:2770-2777.

\bibitem{imb}
Ibragimov NH. Elementary Lie group analysis and ordinary differential equations. New York: John Wiley \& Sons; 1999.

\bibitem{mahomed:1985}
Mahomed FM and Leach PGL. The linear symmetries of a nonlinear differential equation. Quest. Math. 1985;8:241-274.




\bibitem{abraham:1992}
Abraham-Shrauner B, Guo A. Hidden symmetries associated with the projective group of nonlinear first-order ordinary differential equations. J. Phys. A: Math. Gen. 1992;25:5597.

\bibitem{abraham-shrauner:4809}
Abraham-Shrauner B. Hidden symmetries and linearization of the modified Painlev$\acute{e}$-Ince equation. J. Math. Phys. 1993;34:4809-4816.

\bibitem{abraham}
Abraham-Shrauner B, Govinder KS, Leach PGL. Integration of second order ordinary differential equations not possessing Lie point symmetries. Phys. Lett. A 1995;203:169-174.

\bibitem{govinder:1995}
Govinder KS, Leach PGL. On the determination of nonlocal symmetries. J. Phys. A : Math. Gen. 1995;28:5349.

\bibitem{sir1}
Lakshmanan M, Senthilvelan M. Direct integration of generalized Lie or dynamical symmetries of three
degrees of freedom nonlinear Hamiltonian systems: integrability and separability. J. Math. Phys. 1992;33:4068-4077.

\bibitem{hydon}
Hydon PE. Symmetry methods for differential equations : a beginner's guide, Cambridge: Cambridge University Press; 2000.

\bibitem{carinena:2009}
Cari$\tilde{n}$ena JF, Guha P, Ra$\tilde{n}$ada MF. Higher-order Abel equations: Lagrangian formalism, first integrals and Darboux polynomials. Nonlinearity 2009;22:2953-2969.

\bibitem{carinena}
Cari$\tilde{n}$ena JF, Guha P, Ra$\tilde{n}$ada MF. A geometric approach to higher-order Riccati chain: Darboux polynomials and constants of the motion. J. Phys.: Conf. Ser. 2009;75:012009.


\bibitem{chandrasekar:royal1}
Chandrasekar VK, Senthilvelan M, Lakshmanan M. On the complete integrability and linearization of certain second order nonlinear ordinary differential equations. Proc. R. Soc. A. 2005;461:2451-2476.

\bibitem{euler:2009}
Euler M, Leach PGL. Aspects of proper differential sequences of ordinary
differential equations. Theor. Math. Phys. 2009;159:474-487.

\bibitem{euler:2007}
Euler M, Euler N, Leach PGL. The Riccati and Ermakov-Pinney hierarchies. J. Nonlinear Math. Phys. 2007;14:290-310.
  
\bibitem{bruzon}
Bruz$\acute{o}$n MS, Gandarias ML, Senthilvelan M. Nonlocal symmetries of Riccati and Abel chains and their similarity reductions. J. Math. Phys. 2012;53:023512.
  
\bibitem{dar1}
Darboux G. M$\acute{e}$emoire sur les $\acute{e}$quations diff$\acute{e}$rentielles alg$\acute{e}$briques du premier ordre et du premier degr$\acute{e}$. Bull. Sci. Math. 1878;2:60.

\bibitem{pandey:102701}
Pandey SN, Bindu PS, Senthilvelan M, Lakshmanan M. A group theoretical identification of integrable
cases of the Li$\acute{e}$nard type equation $\ddot{x}+f(x)\dot{x}+g(x)=0$: Part II: Equations having maximal Lie point symmetries. J. Math. Phys. 2009;50:102701-25.


\bibitem{chandru}
Gladwin Pradeep R, Chandrasekar VK, Senthilvelan M, Lakshmanan M. Dynamics of a completely integrable $N$-coupled Li\'enard-type nonlinear oscillator. J. Phys. A: Math. Theor. 2009;42:135206.
  

\end{thebibliography}
\end{document}